\documentclass[sigconf]{acmart}
\usepackage{algorithm}
\usepackage{algpseudocode,algorithmicx}
\usepackage{amsmath,amssymb,amsfonts,amsthm,graphicx}
\usepackage{multirow,xspace,multicol,vwcol,forest}
\usepackage{tikz,pgfplots}
\usepackage{multirow}
\usepackage{enumerate}
\usepackage{hyperref}

\newtheorem{claim}{Claim}
\usepackage{xcolor,colortbl}

\def\extend{\ensuremath{\textsc{EXTEND}}}
\def\kextend{\ensuremath{\textsc{K-EXTEND}}}
\def\alg{\ensuremath{\textsc{ALG}}\xspace}
\def\S{\ensuremath{\mathcal{S}}\xspace}

\def\reoptsk{Reopt-CSP\ensuremath{_{\mathcal{M}_{t+k}}}\xspace}
\def\reoptsmk{Reopt-CSP\ensuremath{_{\mathcal{M}_{t-k}}}\xspace}

\def\reoptp{Reopt-CSP\ensuremath{_{\mathcal{M}_{l+1}}}\xspace}
\def\reoptm{Reopt-CSP\ensuremath{_{\mathcal{M}_{l-1}}}\xspace}
\def\reoptpk{Reopt-CSP\ensuremath{_{\mathcal{M}_{l+k}}}\xspace}
\def\reoptmk{Reopt-CSP\ensuremath{_{\mathcal{M}_{l-k}}}\xspace}

\def\alg{\ensuremath{\textsc{Alg}}\xspace}

\def\O2S{{\sc Online $2$-Search}\xspace}

\def\vopt{\ensuremath{{v_{opt}}}\xspace}
\def\vpopt{\ensuremath{{v^\prime_{opt}}}\xspace}
\def\vpsol{\ensuremath{{v^\prime_{sol}}}\xspace}
\def\vsol{\ensuremath{{v_{sol}}}\xspace}

\usepackage{multicol}

\begin{document}
\title{Reoptimization of the Closest Substring Problem under Pattern Length Modification}
\author{Jhoirene B. Clemente	\and Henry N. Adorna}
\affiliation{%
 \institution{University of the Philippines Diliman}
 \department{Department of Computer Science}
 \city{Quezon City}
 \postcode{1101}
 \country{Philippines}
}\email{{jbclemente,hnadorna}@up.edu.ph}

\begin{abstract}
This study investigates whether reoptimization can help in solving the closest substring problem. We are dealing with the following reoptimization scenario. Suppose, we have an optimal $l$-length closest substring of a given set of sequences \S. How can this information be beneficial in obtaining an $(l+k)$-length closest substring for \S ? In this study, we show that the problem is still computationally hard even with $k=1$. We present greedy approximation algorithms that make use of the given information and prove that it has an additive error that grows as the parameter $k$ increases. Furthermore, we present hard instances for each algorithm to show that the computed approximation ratio is tight. We also show that we can slightly improve the running-time of the existing polynomial-time approximation scheme (PTAS) for the original problem through reoptimization.

\end{abstract}

%
%

\ccsdesc[500]{Mathematics of computing~Combinatorial optimization}
\ccsdesc[500]{Mathematics of computing~Approximation algorithms}

\keywords{approximation, reoptimization, closest substring problem}
\maketitle

\section{Introduction}
Given a set of sequences $\mathcal{S} = \{S_1, S_2, \ldots ,S_t \}$ defined over some alphabet $\Sigma$, where each $|S_i|=n$, and for some $l < n$, find a string $v \in \Sigma^l$ and a set containing $\{y_i\}$, where each $y_i$ is a substring of $S_i \in$ \S, such that the total Hamming distance $\sum\limits_{i}^t d(v, y_i)$ is minimized. We call $v$ the $l$-length \emph{ closest substring} of \S. The string $v$ is also called the \emph{consensus} of the set $\{y_1, y_2, \ldots, y_t\}$. Solutions to this problem has been applied to variety of pattern identification ranging from biological sequences to text mining. Not to mention its many application to other discrete structures such as graphs.

The problem of finding the closest substring is NP-hard \cite{Garey1979}, i.e., unless P=NP, there does not exists a polynomial-time exact solution for the problem. Therefore, approaches such as finding near-optimal solutions has been widely used to address the intractability of the problem. {\it Approximation} is one among these approaches. In this approach, algorithms are required to have provable error bounds. Here, one can compute a constant $c$, called the \emph{approximation ratio} which serves as a performance guarantee of an algorithm. A hierarchy exists for NP-hard optimization problems showing that, while others have constant-factor approximation ratio, some still are not even possible to approximate. Examples of inapproximable problems include, the unrestricted traveling salesman problem \cite{Sahni1976} and the maximum subgraph problem \cite{Yannakakis1979}. Therefore, we use another approach that goes hand in hand with approximation, called {\it reoptimization}. 

Reoptimization was first mentioned in \cite{Schaffter1997}. Reoptimization is used to solve computational problems that are defined over instances that change over time. To illustrate the concept, consider a railway system with an optimal routing schedule. As part of development, new stations or connections will be added to the railway system. Thus, as a consequence, a new routing schedule for the new railway system is required. Reoptimization has been applied to similar studies including finding the shortest path in \cite{Nardelli2003}, finding the minimum spanning tree in \cite{Thorup2000} and some of its variants with edge weights in \cite{Ribeiro2007} \cite{Cattaneo2010}. It is also used in providing reoptimization solutions for vehicle routing problem \cite{Secomandi2009}, and the facility location problem \cite{Shachnai2012}.

For some instances, the optimal routing schedule remains to be optimal after the modification, but for some, however trivial the modification, the problem of coming up with a new routing schedule remains to be computationally hard \cite{Bockenhauer2008}. In line with this, several studies investigate the benefit of reoptimization when applied to computationally hard problems.For some problems, the given optimal solution provides a good approximate solution to the new instance. Moreover, it was shown that reoptimization can help to either improve the approximability and even provide a PTAS for some problems that are APX-hard \cite{Bockenhauer2008,Zych2012}. These results include improvements for the metric-traveling salesman problem \cite{Bockenhauer2008}, the Steiner tree problem \cite{Hromkovic2009,Bilo2012,Bockenhauer2012}, the common superstring problem \cite{Bilo2011}, and hereditary graph problems \cite{Boria2012, Boria2012b}.

The first application of reoptimization for the closest substring problem has been shown from our initial 
work in \cite{Clemente2014,Clemente2015}. 
In \cite{Clemente2014}, we proved that CSP obeys a certain property called {\it self-reducibility}. We also proved that all problems that are polynomial-time reducible to a self-reducible problems admits the same property. The simple idea behind this property is that we can easily break-down any given instance of the problem to a smaller instance, such that whenever there exists a solution to a smaller instance, we can easily make it feasible to the larger instance. 

Initial findings in \cite{Clemente2015}, focused on a reoptimization variant characterized by adding a new sequence in \S. In other words, we have an additional information that is the optimal closest substring for a subset of sequences. Furthermore, we also showed that can we obtain an error that grows as the number of additional sequences is increasing. With the same approximation ratio of the PTAS in \cite{Li1999}, we can improve the running time from $O(l(tn)^{r+1})$ to $O(ltn((t-r)n)^r)$. 

In this paper, we will explore the corresponding reoptimization variant of CSP. We will start with the simple case where the pattern length is increased by $1$. Let us define \reoptp as follows. Given a set of sequences $\mathcal{S} = \{S_1, S_2, \ldots, S_t\}$ and an optimal closest substring $v_{opt}$ of length $l$. Find the closest substring of length $l+1$. Later on we generalized the reoptimization variant to \reoptpk. Given the same set of sequences \S, we investigate whether a given optimal $l$-length closest substring will be beneficial or not in finding an $(l+k)$-length closest substring. 

The paper is organized as follows. In Section 2, we showed that even though we have an additional information regarding $l$-length closest substring, solving \reoptp and \reoptpk remains to be computationally hard. In sections 3 and 4, we provide approximation algorithms for \reoptp and \reoptpk respectively. In section 5, we showed how reoptimization can be beneficial in improving the running time of the PTAS for CSP. Lastly, we conclude this paper in section 6.

 

\section{Hardness Result} \label{sec:reoptp_hardness} 
\begin{theorem}
\reoptp is NP-hard.
\label{thm:reoptp_hard}
\end{theorem}

\begin{proof}
Towards contradiction, suppose \reoptp problem is polynomial-time solvable, then there $\exists$ an optimal polynomial-time algorithm \alg for \reoptp. Now, we present an iterative algorithm for closest substring problem utilizing \alg. We will start with a trivial closest substring of length $l =1$. 
For any valid set of sequences, any symbol that is present in all sequences is an optimal solution for \S, except for the trivial case where the set of alphabets in $S_i$'s are disjoint.

Using the optimal closest substring of length $1$, we can obtain an optimal solution of length $2$ in polynomial-time using \alg. Iteratively, we can use the optimal solution of length $i$ to get the optimal solution of length $i+1$, for $2 \leq i \leq l$. Ultimately, we arrive to an optimal solution of an arbitrary length $l$ in polynomial-time. 
However, the closest substring problem is NP-hard. 
Thus, \reoptp must also be NP-hard.
\end{proof}

Using Theorem \ref{thm:reoptp_hard}, we have the following corollary.

\begin{corollary}
\reoptpk is NP-hard.
\end{corollary}

\section{Approximation Algorithms} \label{sec:reoptp_approx} 

It is natural to think that the given optimal solution already provides a good approximate solution for \reoptp. Here, we investigate possible transformations of $v_{opt}$ in order to obtain a feasible solution for \reoptp, as well as the approximation ratio of the best possible solution from transforming $v_{opt}$.

Let $v^\prime_{opt}$ be the optimal $(l+1)$-length closest substring of \S and $OPT = (y_1, \ldots, y_t)$ be the sequence of closest substrings of $v_{opt}$ in \S. In order to obtain a feasible solution of length $l+1$ from a given $l$-length pattern, we define algorithm EXTEND in Algorithm \ref{alg:extend}.

\begin{algorithm}
\begin{algorithmic}[1]
\Procedure{EXTEND}{\vsol, \S}
\State{obtain $SOL = \{y_1, y_2, \ldots, y_t\}$ from \vsol} \Comment{Get the closest substring $y_i$ from \vopt for each $S_i$}

\For{{\bf each} $x \in \{0,1\}^t$ }
\For{ \textbf{each} $y_i \in SOL$ }
\If{$x[i] == 1$}
\State{$y^\prime_i = left(y_i, S_i)$}
\Else 
\State{$y^\prime_i = right(y_i, S_i)$}
\EndIf
\EndFor
\State{$SOL^\prime = \{y^\prime_1, y^\prime_2, \ldots, y^\prime_t \}$}
\EndFor
\State{{\bf return} consensus $v^\prime_{sol}$ with minimum $cost(SOL^\prime)$}
\EndProcedure
\end{algorithmic}

\caption{\textbf{(EXTEND).} Approximation algorithm for transforming any given $l$-length pattern \vsol to obtain a feasible solution $v^\prime_{sol}$ of length $l+1$ for \reoptp.}
\label{alg:extend}
\end{algorithm}

Algorithm \ref{alg:extend} extends each $y_i$ either to the left or to the right to obtain an occurrence of length $l+1$. Let $y^\prime_i$ be the extended substring from $y_i$ of length $l+1$, such that 

\[
cost(v^\prime_{sol}) = \sum\limits_{i=1}^t d(v^\prime_{sol}, y^\prime_i)
\]

is minimized over all combinations of left and right extensions of each $y_i$ in $OPT$, with respect to their consensus substring $v^\prime_{sol}$. Naively, we can get the best solution from \extend(\vopt) in $O(2^t)$. Due to the transformation, the quality of $v^\prime_{sol}$ with respect to a given $v_{opt}$ is

\[
cost(v^\prime_{sol}) = cost(v_{opt}) + t.
\]
Since $cost(v_{opt}) \leq cost(v^\prime_{opt})$ in this type of modification, we have an additive approximation ratio of

\[
cost(v^\prime_{sol}) = cost(v^\prime_{opt}) + t.
\]
From the given computations, we have the following theorem.

\begin{theorem}
Procedure EXTEND in Algorithm \ref{alg:extend} is an approximation algorithm for \reoptp with cost at most $cost(v^\prime_{opt}) + t$ which runs in $O(2^t)$.
\end{theorem}


Though it might seem that the shown approximation ratio for \reoptp is a trivial upper bound, we will show that the ratio is indeed tight by showing a set of hard instances for \reoptp. Let us consider an instance for \reoptp. Let \S be the following set of sequences, 

 \begin{center}
\begin{tabular}{ c c c c c c c c c c}
$S_1$: & $\alpha_1$ & {\bf B} & {\bf B} & $\alpha_1$ & A & B & A \\
$S_2:$ & $\alpha_2$ & {\bf B} & {\bf B} &$\alpha_{2}$ & A & A & A \\
$S_3:$ & $\alpha_3$ & {\bf B} & {\bf B} &$\alpha_{3}$ & A & A & A \\
: & & & \ldots & & & \\
$S_t:$ & $\alpha_t$ & {\bf B} & {\bf B} & $\alpha_t$ & A & A & A \\
\end{tabular}
\end{center}

where $A, B, \alpha_i \in \Sigma$. The optimal $2$-length closest substring of \S is $``BB"$ with $cost(``BB") = 0$. However, all possible extension of $``BB"$ will incur an additional cost of $t$. On the other hand, a suboptimal solution $``AA"$ could have been a better option when transformed to $``AAA"$. This particular example can be generalized to a set of input instances for \reoptp. The description of such instances is described in the proof of the following claim. 

\begin{claim}
There exists an instance \S and a given $v_{opt}$ for \reoptp such that the $$cost(\extend(\vopt)) = cost(v^\prime_{opt}) + (t-1).$$
\end{claim}

\begin{proof} 
We prove the following claim by describing a set instances for \reoptp. Let \S $= \{S_1, S_2, \ldots S_t\}$ be the set of sequences defined over the alphabet $\Sigma$ containing the subset of symbols $\{A, B, \alpha_1, \ldots, \alpha_t\}$. The set \S is defined such that $\exists \ j \in \{1,\ldots,t\}$, where $S_j$ is of the following form

 \[
 S_j = \alpha_j \underbrace{B \ldots B}_l \alpha_{j} \underbrace{A \ldots B}_l A,
 \] 
 and all the remaining sequences in \S is described as follows 
 \[
 S_i = \alpha_i \underbrace{B \ldots B}_l \alpha_{i} \underbrace{A \ldots A}_l A.
 \]
 
For illustration purposes, we have the following alignment.
 \begin{center}
 
\begin{tabular}{ c c c c c c c c c c c c c c}
$S_1$: & $\alpha_1$ & {\bf B}& \ldots & {\bf B} & $\alpha_1$ & A & \ldots &A& A & A \\
$S_2:$ & $\alpha_2$ & {\bf B} & \ldots & {\bf B} &$\alpha_{2}$ & A & \ldots &A& A & A \\
$S_3:$ & $\alpha_3$ & {\bf B} & \ldots & {\bf B} &$\alpha_{3}$ & A & \ldots &A& A & A \\
 : & & & \ldots & & & \\
$S_j:$ & $\alpha_j$ & {\bf B} & \ldots & {\bf B} &$\alpha_{j}$ & A & \ldots &A & B & A \\
 : & & & \ldots & & & \\
$S_t:$ & $\alpha_t$ & {\bf B} & \ldots & {\bf B} & $\alpha_t$ & A & \ldots &A& A & A \\
\end{tabular}
\end{center}

The optimal solution of length $l$ is the closest substring $B^l$ with $cost(B^l) = 0$. However, the best possible solution from $B^l$ will incur an additional cost of $t$, i.e., $cost(\extend(B^l)) = cost(B^l) + t$. On the other hand, a suboptimal solution $A^l$, with $cost(A^l) = 1$, can be transformed into the optimal solution $A^{l+1}$, i.e., $\extend(A^l) =A^{l+1}$, with $$cost(\extend(A^l)) = 1.$$ Therefore, showing 

\[
cost(\extend(v_{opt})) = cost(v_{opt}) + t,
\]
\[
 cost(v_{opt}) = cost(v^\prime_{opt}) - 1,
\]
\[
cost(\extend(v_{opt})) = cost(v^\prime_{opt}) + t-1.
\]
\[
cost(v^\prime_{sol}) = cost(v^\prime_{opt}) + t-1.
\]
\end{proof}

%

\begin{theorem}
\label{thm:sigma_extend}
If there exists a $\sigma$-approximation algorithm for CSP, then there exists an approximation algorithm with ratio $$\frac{(2 \sigma -1)}{\sigma} cost(v^\prime_{opt}) + \frac{ (\sigma-1) }{\sigma} (t-1) $$ for \reoptp.
\end{theorem}

\begin{proof}
Let \alg return the minimum between the results of $ALG_1$ and $ALG_2$. Let $ALG_1$ be the reoptimization algorithm \extend \ and $ALG_2$ be a existing $\sigma$-approximation algorithm. We have the following computation of $cost(\vpsol)$.\\

\begin{tabular}{l l l}
$cost(\vpsol_{_A}) $ & $\leq$ & $cost(\vpopt) + t-1$ \\
$cost(\vpsol_{_B})$ & $\leq$ & $\sigma \cdot cost(\vpopt)$ \\ \ \\
$cost(\vpsol)$ & $=$ & $min\{cost(SOL_A), cost(SOL_B)\}$\\
 & $\leq$ & $\frac{(\sigma-1)( cost(\vpopt) + t-1) + \sigma cost(\vpopt)}{ \sigma}$\\
 & $\leq$ & $\frac{(\sigma-1)cost(\vpopt) + (\sigma-1)(t-1) + \sigma cost(\vpopt)}{ \sigma}$\\
 & $\leq$ & $\frac{(2\sigma-1) cost(\vpopt) + (\sigma-1)(t-1) }{ \sigma}$\\
 & $\leq$ & $\frac{(2\sigma-1) }{ \sigma}cost(\vpopt)+ \frac{ (\sigma-1) }{\sigma} (t-1)$\\
\end{tabular}\\
\end{proof}
In the following corollary, we identify properties of some input instances where we can actually benefit from the additional information in \reoptp. 

\begin{corollary}\label{cor:sigma_reopt}
If $(t-1) < (\sigma -1) cost(\vpopt)$ for some feasible instance $\S^\prime$ , then algorithm EXTEND for \reoptp is an advantage over any existing $\sigma$-approximation algorithm for CSP.
\end{corollary}

On the contrary, if $(t-1) \geq cost(\vpopt)$, it is better to solve $\S^\prime$ from scratch using the existing $\sigma$-approximation algorithm. In this case, the given optimal solution is not beneficial in improving the quality of the solution.

\section{Generalization} \label{sec:reoptp_generalization} 
The procedure EXTEND in Algorithm \ref{alg:extend} can be generalized to obtain a feasible solution of length $l+k$. We illustrate all possible $k$ extensions of a sample substring as follows. Consider a subtring $y_i = DEF$ in $S_i$. For $k =2$, we have $3$ possible values for $y^\prime_i$. For the rest of our discussion, we may refer to the additional substrings in $y^\prime_i$ as the $k$ flanking substrings of $y_i$ in $y^\prime_i$.
 
\begin{figure}[h]
\centering
\begin{tabular}{ c l c c c c c c c c c c c }
$S_i$ & $:$ & A & B & C & {\bf D} & {\bf E} & {\bf F} & G & H & I & J \\
$y_i $& $:$ & & & & {\bf D} & {\bf E} & {\bf F} & & & & \\ \ \\
$y^\prime_i $& $:$ & & B & C & {\bf D} & {\bf E} & {\bf F} & & & & \\
& & & & C & {\bf D} & {\bf E} & {\bf F} & G & & & \\
& & & & & {\bf D} & {\bf E} & {\bf F} & G & H & & \\
\end{tabular}
\label{fig:k-extend}
\end{figure}

A substring $y_i$ of length $l$ can be extended to at most $k+1$ possible $y^\prime_i$ in $S_i$. Procedure K-EXTEND in Algorithm \ref{alg:k_extend} works by getting all possible combination of extensions from the left and right of each occurrence.

\begin{algorithm}

\begin{algorithmic}[1]
\Procedure{K-EXTEND}{\vsol, \S}
\State{obtain $SOL = \{y_1, y_2, \ldots, y_t\}$ from \vsol} \Comment{Get the closest substring $y_i$ from \vopt for each $S_i$}
\For{\textbf{each} $x \in \{0, 1,\ldots,k\}^t$}
	\For{\textbf{each} $y_i \in SOL$}
		\State{$\text{\it start} = \text{starting position of } y[i]$ in $S_i$}
		\State{$\text{\it start} = \text{\it start} + x[i]$}
		\State{$y^\prime = S_i[\text{\it start}+0 : \text{\it start} +k]$}
	\EndFor
\EndFor
\EndProcedure
\end{algorithmic} 

\caption{\textbf{(K-EXTEND).} Generalization of the EXTEND algorithm for \reoptpk}
\label{alg:k_extend}
\end{algorithm}

\begin{theorem}
Procedure K-EXTEND in Algorithm \ref{alg:k_extend} is an approximation algorithm for \reoptpk with cost at most $cost(\vpopt) + kt $ which runs in $O(t \cdot k^t)$.
\end{theorem}

\begin{proof}
Exhausting all possible extensions will take $O((k+1)^t)$ steps. Extracting substrings from each sequence will take $O(t)$ steps. Therefore, Algorithm \ref{alg:k_extend} has a worst case time complexity of $O(t\cdot k^t)$. 

For the approximation ratio of K-EXTEND, we use the proof of Theorem \ref{thm:sigma_extend} to give us an upper bound of $cost(\vpsol) = cost(\vpopt) + kt$ for \reoptpk.

\end{proof}

\section{Improving the PTAS} \label{sec:reoptp_ptas} 

Recall the sampling-based PTAS in \cite{Li1999}. For each parameter $r$, it describes an approximation algorithm for CSP that outputs a solution $\vsol$ with $$cost(\vsol) \leq \left( 1 + \frac{4|\Sigma| - 4}{\sqrt{e}\sqrt{4r+1}-3} \right) \cdot cost(\vopt)$$ in $O(l(tn)^{r+1})$ time. In this section, we will show that it is also possible to adapt the general idea of the existing PTAS from \cite{Li1999} for improving the approximation ratio of K-EXTEND algorithm. Moreover, we argue that we also improve the running time of the existing PTAS for \reoptpk. 

Note that, by exhausting all possible substring alignments in \S, we can get the optimal closest substring in $O((tn)^t)$. The PTAS from Li et. al. \cite{Li1999} explores a subset of this search space by limiting the number of substrings in the alignments. Instead of exhausting all possible alignments of $t$ substrings in \S, the PTAS explores all possible alignments of $r$ substrings present in \S, where parameter $r \leq t$. For some fix $r$, it is easy to see how the problem admits a polynomial-time approximation solution in $O((tn)^r)$. 
 
Before we proceed with the discussion of how we aim to improve the PTAS via reoptimization. Let us present the following concepts. An $r$-\textit{sample} from a given instance $\mathcal{S}$ or a set of sequences, i.e., 
$$r\text{-}sample(\mathcal{S})\ = \{y_{i_1}, y_{i_2}, \ldots, y_{i_r}\},$$ is a collection of $r$ $l$-length substrings from $\mathcal{S}$. Repetition of substrings are allowed for as long as no two substrings are obtained from the same sequence. Let $R(\mathcal{S})$ denote the set of all possible $r$-\textit{sample} from $\mathcal{S}$. The total number of 
samples in $\mathcal{S}$ is ${tn \choose r}$ which is bounded above by $O((tn)^r)$. Note that, a consensus pattern is polynomial-time computable from a given $r$-sample. This is done by simply getting the column-wise majority symbol from an alignment of a given set of equal length substrings. 

We present an approximation algorithm for \reoptpk in Algorithm \ref{alg:iptas2}. The algorithm outputs the best between two feasible solutions $\vpsol_{_A}$ and$\vpsol_{_B}$. The first solution $\vpsol_{_A}$ is obtained from using K-EXTEND on the given optimal solution \vopt. The second feasible solution is obtained by minimizing the cost among all $r$-samples obtained from the set $\{ R(\S) \setminus R(SOL^\prime_A)\}$, where $R(SOL^\prime_A)$ contains the set of occurrences of $\vpsol_{_A}$ from K-EXTEND Algorithm.

\begin{algorithm}[h]
\caption{Modification of PTAS from \cite{Li1999} for \reoptpk. Given a set of sequences \S$ = \{S_1, S_2, \ldots, S_t\}$ and a corresponding optimal $l$-length closest substring \vopt, algorithm outputs a feasible $(l+k)$-length closest substring $v^\prime_{sol}$}
\label{alg:iptas2}

\begin{algorithmic}[1]
\State{$\vpsol_{_A} = \text{\kextend(\vopt)}$} 
\State{$\vpsol_{_B} = \emptyset $}
\State{$min= \infty$}
\For{\textbf{each} $(l+k)$-length $r$-$samples \ \{y_{i_1}, y_{i_2}, \ldots, y_{i_r} \} \in \{ R(\S) \setminus R(SOL^\prime_A)\}$ }
\State{$\vsol=$ K-BEST-ALIGN$(\{y_{i_1}, y_{i_2}, \ldots, y_{i_r} \} , \S)$}

\If{$cost(\vsol) < min$ }
\State{$min = cost(\vsol)$}
\State{$\vpsol_{_B} = \vsol$}
\EndIf
\EndFor 
\State{$\vpsol = min(\vpsol_{_A}, \vpsol_{_B})$}

\State{{\bf return} \vpsol}
\end{algorithmic} 
\end{algorithm}
We argue that the algorithm can actually skip a portion of the sample search space. Thereby, maintaining the same approximation ratio while improving the running time. To illustrate the idea of the algorithm, we abstracted the sample space using the following figure. 
\begin{figure}[h]
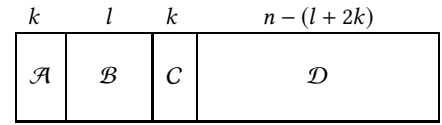

\centering
\begin{tabular}{lcllccllll}
$k$ & \multicolumn{1}{l}{} & $l$ & & \multicolumn{1}{l}{$k$} & \multicolumn{1}{l}{} & & $n - (l +2k)$ & & \\ \hline
\multicolumn{1}{|l|}{\multirow{3}{*}{$\mathcal{A}$}} & \multicolumn{3}{c|}{\multirow{3}{*}{$\mathcal{B}$}} & \multicolumn{1}{c|}{\multirow{3}{*}{$\mathcal{C}$}} & \multicolumn{5}{c|}{\multirow{3}{*}{$\mathcal{D}$}} \\
\multicolumn{1}{|l|}{} & \multicolumn{3}{c|}{} & \multicolumn{1}{c|}{} & \multicolumn{5}{c|}{} \\
\multicolumn{1}{|l|}{} & \multicolumn{3}{c|}{} & \multicolumn{1}{c|}{} & \multicolumn{5}{c|}{} \\ \hline
\end{tabular}
\caption{Abstraction of the search space in $R(\S)$, where $R(\S) =\mathcal{A} \cup \mathcal{B} \cup \mathcal{C} \cup \mathcal{D}$. The substring $\vpsol_{_A}$ is the best solution obtained from exhausting regions $\mathcal{A} \cup \mathcal{B} \cup \mathcal{C}$, while $\vpsol_{_B}$ is obtained by exhausting region $\mathcal{D}$.}
\label{fig:sample_region}
\end{figure}

Suppose each sequence in \S has a uniform length of $n = m + l+ 2k$. Let us partition the set of sample spaces into $4$ regions as shown in Figure \ref{fig:sample_region}. Recall the definition of an $r$-sample in the previous section. Region $\mathcal{B}$ consists of the set of all $r$-samples obtained from the occurrences of \vopt. Regions $\mathcal{A}$, $\mathcal{B}$, and $\mathcal{C}$ consist of the set of $r$-samples obtained from the occurrences of \vopt including $k$ flanking substrings to the left and right of each occurrence. We have the following illustration to visualize the set of substrings where samples from regions $\mathcal{A}$, $\mathcal{B}$, and $\mathcal{C}$ were taken from.
\begin{figure}[h]
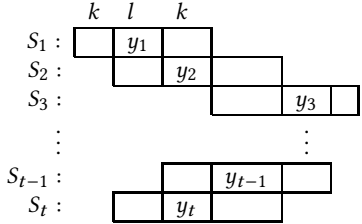

\begin{center}
\begin{tabular}{rccccccccccc}
 & \multicolumn{1}{l}{$k$} & \multicolumn{1}{l}{$l$} & \multicolumn{1}{l}{$k$} & \multicolumn{1}{l}{} & \multicolumn{1}{l}{} & \multicolumn{1}{l}{} & \multicolumn{1}{l}{} & \multicolumn{1}{l}{} & \multicolumn{1}{l}{} & \multicolumn{1}{l}{} & \multicolumn{1}{l}{} \\ \cline{2-4}
\multicolumn{1}{r|}{$S_1:$} & \multicolumn{1}{c|}{} & \multicolumn{1}{c|}{$y_1$} & \multicolumn{1}{c|}{} & & & & & & & & \\ \cline{2-5}
$S_2:$ & \multicolumn{1}{c|}{} & \multicolumn{1}{c|}{} & \multicolumn{1}{c|}{$y_2$} & \multicolumn{1}{c|}{} & & & & & & & \\ \cline{3-7}
$S_3:$ & & & \multicolumn{1}{c|}{} & \multicolumn{1}{c|}{} & \multicolumn{1}{c|}{$y_3$} & \multicolumn{1}{c|}{} & & & & & \\ \cline{5-7}
$\vdots$ & & & & & $\vdots$ & & & & & & \\ \cline{4-6}
$S_{t-1}:$ & \multicolumn{1}{l}{} & \multicolumn{1}{l|}{} & \multicolumn{1}{l|}{} & \multicolumn{1}{l|}{$y_{t-1}$} & \multicolumn{1}{l|}{} & \multicolumn{1}{l}{} & \multicolumn{1}{l}{} & \multicolumn{1}{l}{} & \multicolumn{1}{l}{} & \multicolumn{1}{l}{} & \multicolumn{1}{l}{} \\ \cline{3-6}
$S_t:$ & \multicolumn{1}{c|}{} & \multicolumn{1}{c|}{} & \multicolumn{1}{c|}{$y_t$} & \multicolumn{1}{c|}{} & & & & & & & \\ \cline{3-5}
\end{tabular}
\caption{Occurrences of \vopt in each $S_i$. Each occurrence $y_i$ has a left and a right $k$ flanking substring. Samples that are taken from these substrings are covered in regions $\mathcal{A}$, $\mathcal{B}$, and $\mathcal{C}$ in Figure \ref{fig:sample_region}. }
\label{fig:k_flanking}
\end{center}
\end{figure}

The above illustration captures the fact that occurrences of \vopt in \S may not necessarily align in terms of their starting position. Without lost of generality, we assume that $k$ flanking substrings on the left and right of each occurrence exist in \S. The remaining parts of \S that is not considered in Figure \ref{fig:k_flanking} comprises the samples in region $\mathcal{D}$. 

\begin{theorem}
Algorithm \ref{alg:iptas2} is a $1 + \frac{4|\Sigma| - 4}{\sqrt{e} (\sqrt{4r+1}-3)} $ approximation algorithm for \reoptpk which runs in $O(ltn \cdot (tn)^r - t^r)$.
\end{theorem}

\begin{proof}
Algorithm 7 uses K-EXTEND which runs in $O(t \cdot k^t)$. The sampling step in lines 4-10 runs in $O((tn)^r - (t^r (l+2k)^r))$. For small values of $k$, we have a total running time of $O((tn)^r - (tl)^r)$. However, for large values of $k$, the algorithm will be dominated by the running time of the K-EXTEND which is $O(t \cdot k^t)$. 

An algorithm has to cover all possible $r$-samples in \S in order to achieve the desired competitive ratio as the PTAS. This is equivalent to covering all samples that are obtained from regions $\mathcal{A}$ to $\mathcal{D}$ in Figure \ref{fig:sample_region}. The K-EXTEND algorithm already covers regions $\mathcal{A}$, $\mathcal{B}$, and $\mathcal{C}$. Due to the exhaustiveness of K-EXTEND, the feasible solution $\vpsol_{_A}$ has the local minimum cost when samples obtained from $\mathcal{A}$- $\mathcal{C}$ are considered. The remaining space that is not covered by the K-EXTEND is handled by the sampling based approach in lines 4 to 10 of the Algorithm \ref{alg:iptas2}. 
Thus, maintaining the same approximation ratio of PTAS in \cite{Li1999}.
\end{proof}

We can see in this scenario, how the amount of information is useful in Algorithm \ref{alg:iptas2}. As we have more information about the optimal solution, or equivalently, if we have an optimal solution for a longer sequence, i.e., we have a smaller value of $k$, then we can actually get an advantage over the existing PTAS. But if we have little information, i.e., larger value of $k$, it is advisable to solve the problem from scratch, as it is much more expensive to start from \vopt to obtain a solution of longer lengths through K-EXTEND algorithm. 

Reoptimization in this case is helpful and it scales up as $k$ decreases. The observation in \reoptpk is analogous to our result in the previous section for \reoptsk.
 

\section{Decreasing the Pattern Length} \label{sec:reoptp_inverse} 
We study the reoptimization variant of CSP when the pattern length $l$ is increased. In this section, we will investigate the case where we look for smaller pattern length. Let us make use of \reoptm and \reoptmk to denote the case where the pattern length is decreased by $1$ and $k$, respectively. It is natural to think that \reoptmk and is easier than \reoptpk, i.e., we can always get the smaller closest substring (\vpopt) inside the longer closest substrings (\vopt). This is true for some cases. However, we will show some instances where the smaller closest substring is totally different from the longer substring. Let $\mathcal{S} = 
\{ S_1, S_2, \ldots, S_t\}$, where $cost(\vopt) > t$.



\begin{center}
{\setlength{\tabcolsep}{0.5em}}
\begin{tabular}{ r c c c c c c c c}
$S_1$: & {\bf A} & {\bf A} & {\bf A} & B & B & A \\
$S_2:$ & {\bf A} & {\bf A} & {\bf A} & B & B & A \\
$S_3:$ & {\bf A} & {\bf A} & {\bf A} & B & B & A \\
$S_4:$ & {\bf A} & {\bf A} & {\bf A} & B & B & A \\ 
: & & & \ldots & & & \\
$S_t:$ & {\bf B} & {\bf B} & {\bf B} & B & B & A \\
\end{tabular}
\end{center}

In the given instance, we can observe that \vopt$= ``AAA"$ and \vpopt$= ``BB"$ can be totally different in terms of their edit distance even for binary alphabets, i.e., \S is defined over the alphabet $ \Sigma = \{A,B\}$. The optimal $3$-length closest substring has cost equal to $3$. Meanwhile, the optimal $2$-length substring has cost equal to $0$. The the occurrences of the optimal closest substring of smaller length is not necessarily contained in the occurrences of longer closest substring, which can be observed even if the pattern length is decreased by $1$. We can generalize the description to an arbitrary length $l+k$. For $n \geq 2l+k$ and $l \leq 2t$, we can always describe an instance \S such that we cannot transform the given solution to obtain a modified solution for \reoptmk. In such instances, the given optimal $l+k$-length closest substring has cost equal to $l+k$ and an optimal $l$-length closest substring with cost equal to $0$. 

The relationship of \reoptpk and \reoptmk is different as compared with the first reoptimization variant that we studied in \cite{Clemente2015}. In \reoptsk and \reoptsmk, the hardest instance for \reoptsk remains to be the hardest for \reoptsmk in terms of approximability, whereas in reoptimization variants where the pattern length is involved, the hard instance for \reoptmk is not easily realizable from the hard instance of  \reoptpk.


\section{Conclusion}
In this paper, we showed that the reoptimization variant of CSP under pattern length modifications for both the simple case and its generalization are NP-hard. We presented simple greedy algorithms called EXTEND and K-EXTEND in Algorithms \ref{alg:extend} and \ref{alg:k_extend}, respectively. We used a simple idea where the algorithms transform the given optimal solution to become feasible for the instance with longer closest substring length. The running time of algorithm EXTEND is exponential in $t$ with solutions that has a worst case approximation ratio of $cost(\vpopt) +t$. Furthermore, we present a set of hard instances for EXTEND to show that the approximation ratio that we computed is tight. The scenario happens only when the cardinality of the alphabet exceeds $t+2$. We isolated the case where we can actually have an advantage over any existing $\sigma$-approximation algorithm. As a corollary, we showed that we can benefit from K-EXTEND if $(t-1) < (\sigma -1) cost(\vpopt)$, for any existing $\sigma$-approximation algorithm for CSP. We also presented an analogous result from our previous work in \cite{Clemente2015} regarding the running time improvement over the existing PTAS in \cite{Li1999}. Here, we showed that we can maintain the same approximation ratio while saving $O(t^r)$ running time for \reoptpk. For value of parameter $t > rn$, reoptimization variant \reoptpk can be more beneficial for CSP compared to \reoptsk. 

%

\bibliography{library}

\begin{thebibliography}{10}

\bibitem{Bilo2011}
Davide Bil\`{o}, Hans-Joachim B\"{o}ckenhauer, Dennis Komm, Richard
  Kr\'{a}lovi\v{c}, Tobias M\"{o}mke, Sebastian Seibert, and Anna Zych.
\newblock {Reoptimization of the shortest common superstring problem}.
\newblock {\em Algorithmica (New York)}, 61(2):227--251, 2011.

\bibitem{Bilo2012}
Davide Bilo and Anna Zych.
\newblock {New Advances in Reoptimizing the Minimum Steiner Tree Problem}.
\newblock {\em In Proc. of the Mathematical Foundations of Computer Science,
  LNCS}, 7464:184--197, 2012.

\bibitem{Bockenhauer2012}
Hans-Joachim B\"{o}ckenhauer, Karin Freiermuth, Juraj Hromkovi\v{c}, Tobias
  M\"{o}mke, Andreas Sprock, and Bj\"{o}rn Steffen.
\newblock {Steiner tree reoptimization in graphs with sharpened triangle
  inequality}.
\newblock {\em Journal of Discrete Algorithms}, 11(1):73--86, February 2012.

\bibitem{Boria2012b}
Nicolas Boria, J~Monnot, and VT~Paschos.
\newblock {Reoptimization of maximum weight induced hereditary subgraph
  problems}.
\newblock {\em Theoretical Computer Science}, pages 1--12, 2012.

\bibitem{Boria2012}
Nicolas Boria, J\'{e}r\^{o}me Monnot, Vangelis~Th Paschos, Davide Bil\`{o},
  Peter Widmayer, and Anna Zych.
\newblock {Reoptimization of the Maximum Weighted Pk-Free Subgraph Problem
  under Vertex Insertion}.
\newblock 5426:76--87, January 2012.

\bibitem{Cattaneo2010}
Guiseppe Cattaneo, Pompeo Faruolo, Umberto~Ferraro Petrillo, and Guiseppe
  Italiano.
\newblock {Maintaining dynamic minimum spanning trees: An experimental study}.
\newblock {\em Discrete Applied Mathematics}, 158(5):404--425, 2010.

\bibitem{Clemente2014}
Jhoirene Clemente, Jeffrey Aborot, and Henry Adorna.
\newblock {Reoptimization of Motif Finding Problem}.
\newblock In {\em Proc. of the International MultiConference of Engineers and
  Computer Scientists}, volume~I, pages 106--111, 2014.

\bibitem{Clemente2015}
Jhoirene Clemente, Jeffrey Aborot, and Henry Adorna.
\newblock On self-reducibility and reoptimization of the closest substring
  problem.
\newblock {\em Philippine Computing Journal}, volume 10(2):1--7, 2016.

\bibitem{Garey1979}
Michael~R. Garey and David~S. Johnson.
\newblock {\em Computers and Intractability: A Guide to the Theory of
  NP-Completeness}.
\newblock W. H. Freeman \& Co., New York, NY, USA, 1979.

\bibitem{Bockenhauer2008}
Juraj Hromkovi\v{c} Tobias M\"{o}mke Peter~Widmayer {Hans-joachim
  B\"{o}ckenhauer}.
\newblock {On the hardness of reoptimization}.
\newblock {\em In Proc. of the 34th International Conference on Current Trends
  in Theory and Practice of Computer Science (SOFSEM 2008), LNCS}, 4910, 2008.

\bibitem{Hromkovic2009}
Juraj Hromkovi\v{c}.
\newblock {\em Algorithmic Adventures: From Knowledge to Magic}.
\newblock Springer Publishing Company, Incorporated, 1st edition, 2009.

\bibitem{Li1999}
Ming Li, Bin Ma, and Lusheng Wang.
\newblock {Finding similar regions in many strings}.
\newblock {\em Proceedings of the thirty-first annual ACM symposium on Theory
  of computing - STOC '99}, 65(1):473--482, August 1999.

\bibitem{Nardelli2003}
Enrico Nardelli, Guido Proietti, and Peter Widmayer.
\newblock {Swapping a Failing Edge of a Single Source Shortest Paths Tree Is
  Good and Fast}.
\newblock {\em Algorithmica}, pages 56--74, 2003.

\bibitem{Ribeiro2007}
Celso Ribeiro and Rodrigo Toso.
\newblock {Experimental analysis of algorithms for updating minimum spanning
  trees on graphs subject to changes on edge weights}.
\newblock {\em Experimental Algorithms}, 2007.

\bibitem{Sahni1976}
Sartaj Sahni and Teofilo Gonzalez.
\newblock P-complete approximation problems.
\newblock {\em J. ACM}, 23(3):555--565, July 1976.

\bibitem{Schaffter1997}
Markus~W. SchÃ€ffter.
\newblock Scheduling with forbidden sets.
\newblock {\em Discrete Applied Mathematics}, 72(1):155 -- 166, 1997.

\bibitem{Secomandi2009}
N.~Secomandi and F.~Margot.
\newblock {Reoptimization Approaches for the Vehicle-Routing Problem with
  Stochastic Demands}, 2009.

\bibitem{Shachnai2012}
Hadas Shachnai, Gal Tamir, and Tami Tamir.
\newblock {A Theory and Algorithms for Combinatorial Reoptimization}.
\newblock {\em Lecture Notes in Computer Science}, 7256(1574):618--630, 2012.

\bibitem{Thorup2000}
Mikkel Thorup.
\newblock {Dynamic Graph Algorithms with Applications}.
\newblock {\em Proceedings of the 7th Scandinavian Workshop on Algorithm
  Theory}, pages 1--9, July 2000.

\bibitem{Yannakakis1979}
Mihalis Yannakakis.
\newblock The effect of a connectivity requirement on the complexity of maximum
  subgraph problems.
\newblock {\em Journal of the ACM}, 26(4):618--630, oct 1979.

\bibitem{Zych2012}
Anna Zych.
\newblock {\em {Reoptimization of NP-hard Problems}}.
\newblock PhD thesis, ETH Zurich, 2012.

\end{thebibliography}
\bibliographystyle{plain}
\end{document}